# *Splendeurs et misères de* theory of laser beam propagation in nonlinear media

V. V. Semak[1] and M. N. Shneider[2]
[1] Applied Research Laboratory, the Pennsylvania State University, PA
[2] Department of Mechanical and Aerospace Engineering, Princeton University, NJ

E-mail: vxs21@psu.edu and m.n.shneider@gmail.com

*First they ignore you, then they laugh at you, then they fight you, then you win.*
Mahatma Gandhi

**Abstract**
It is demonstrated that current theoretical models utilize equations for description of laser beam propagation in nonlinear media that were deduced under the assumption of homogeneity of dielectric constant of the media and for the case of planar wave front. Here, we deduce complete wave propagation equation that includes inhomogeneity of the dielectric constant and present this propagation equation in compact vector form. Although similar equations are known in the narrow fields, such as, radio wave propagation in ionosphere and electro-magnetic and acoustic wave propagation in stratified media, we develop here a novel approach of using such equations in modeling of laser beam propagation in nonlinear media. The inadequacy of the assumptions under which the propagation equations are derived in the current model is demonstrated. Also, mathematical derivation is presented that describes foundation and validates our previously reported method for modeling of laser beam propagation in nonlinear media by blending solution of linear Helmholtz equation with a correction term representing nonlinear field perturbation expressed in terms of paraxial ray-optics (eikonal) equation. Unlike all previous theoretical approaches developed during past five decades, our approach satisfies correspondence principle since in the limit of zero-length wavelength it reduces from physical to geometrical optics.

## 1. Motivation for this research

The science of nonlinear optics is one of rapidly growing fields driven by multiple important technological applications. Since the invention of lasers this field experienced revolutionary progress fueled by splendid experimental and theoretical results that provided deep understanding of nonlinear response of matter to high intensity electromagnetic wave. However, as we will demonstrate here, this splendor was accompanied by utter and miserable failure in one particular sub-discipline – propagation of laser beam in nonlinear media.

Large number of theoretical works describing laser propagation in nonlinear media was published since the concept of laser beam self-focusing and self-trapping was proposed [1]. After more than 50 years of intense research the theoretical concepts and models of self-focusing, beam self-trapping, filamentation, and filament plasma defocusing and the corresponding mathematical models were formulated. The books and extensive reviews (for example, [2-8]) written within past two decades devote chapters to the detailed description of the peculiarities of this physical phenomenon. Thus, the laser beam propagation in nonlinear media appears as a well-established field of science. Students and researchers who enter the arena of nonlinear optics use this material as “textbook” science and current frontier research of laser beam propagation in nonlinear media deals with ingenious formulations and creative solutions of nonlinear Schrodinger equation that describes laser beam collapse, self-trapping, dispersion, filamentation, modulation instability, pulse splitting, and other extraordinary particularities of nonlinear beam propagation [6,9,10]. Therefore, understandably, it was unexpected to discover that the results obtained with our recent straightforward theoretical model and numerical simulation of ultrahigh intensity laser pulse propagation in gases [11,12] revealed grave flaws in the part of this established science that describes self-focusing, beam self-trapping, filamentation, and continuum generation. In particular, our theoretical consideration showed that the concept of “critical power” of laser beam required for beam self-trapping and filamentation is flawed, the waveguide-like description of beam self-trapping is inadequate, the balance between divergence of the beam due to diffraction and convergence induced by nonlinearity of the medium does not occur, the self-similar solution of propagation equation is non-physical, and use of nonlinear Schrodinger equation gives inadequate description of laser beam propagation in nonlinear media.

Being unable to find errors or invalidating assumptions in our forthright approach we proposed detailed investigation of this discrepancy. The submitted proposal received review from someone “*whose expertise is acknowledged throughout the world of ultrashort laser pulse basic research*”. The anonymous reviewer stated that “… *the authors may have been in a time capsule for the past 2 decades and had missed the very substantial developments in high power femtosecond pulse filamentation*.” Intriguingly, the reviewer agreed with our criticism of the concept of critical power of self-focusing making remarkably astonishing statement that contradicts all publically available information: “… *a rather naïve and oversimplified estimate for the critical power for self-focusing made in the beginning of this field in the early sixties is not correct. I agree as does the entire community of researchers who have worked in this area since*”. This admission is surprising because the concept of critical power is a foundational concept of the “textbook” theory of self-focusing and self-trapping that predicts kilometers long plasma filaments produced in air by a laser and, previously, was used as the justification for established large “Teramobile” research program (see http://www.teramobile.org).

This review from a world renowned researcher as well as the reviews from several journals that rejected our submissions demonstrated superficial knowledge of physics, absence of constructive criticism and, instead, contained emotional attacks, illogical statements such as, “…*self-channeling, which, by definition, occurs above the critical power*” and misleading nomenclature presenting current obviously classical propagation theory as a quantum mechanical theory: “…*the current state of the art in theory and modeling is addressing the rather naïve phenomenological models of Kerr self-focusing and ionization with more rigorous quantum mechanical approaches*”. These discrepancies, inaccuracies and lack of scientific objections strengthen our motivation to prove validity of our works and, therefore, we equipped our time capsule with a warp drive and journeyed back to the source – Maxwell’s equations, in order to review the foundational scientific principles. Our examination promptly exposed the original fundamental flaw that lead to the cascading failures of all current models of laser beam propagation in nonlinear media. Also, this examination provided theoretical justification and proof of validity our previously published approach [11,12]. Below is the account of the results of this journey.

## 2. Formulation of general equation for beam of EMW propagation in NL media

As is well known, the electric and magnetic fields in dielectric media are described by the macroscopic Maxwell's equations:

$$\nabla \bullet \vec{D} = 0, \tag{1}$$

$$\nabla \bullet \vec{B} = 0, \tag{2}$$

$$\nabla \times \vec{E} = -\frac{\partial \vec{B}}{\partial t}, \tag{3}$$

$$\nabla \times \vec{H} = \frac{\partial \vec{D}}{\partial t}, \tag{4}$$

where $\vec{E}$ and $\vec{B}$ are the electric and the magnetic fields, correspondingly , while $\vec{D}$ and $\vec{H}$ are, respectively, the displacement and magnetization fields. The latter fields, sometimes called "macroscopic" fields, reflect the effect from matter and are defined using phenomenological constituent equations that relate them to the "microscopic" electric field, $\vec{E}$, and the magnetic field, $\vec{B}$:

$$\vec{D} = \varepsilon_0 \varepsilon \vec{E}, \tag{5}$$

$$\vec{B} = \mu_0 \mu \vec{H}, \tag{6}$$

where $\varepsilon_0$, $\mu_0$ are the permittivity and the permeability of free space and $\varepsilon$, $\mu$ are the permittivity and the permeability of material.

Following known procedure we take *curl* vector operator of both sides of equation (3) and time derivative of both sides of equation (4) and assume that the magnetic effect of media is negligible, i.e. $\mu = 1$. Then, using equations (5) and (6), we can eliminate magnetic and magnetization fields obtaining equation for the electric field

$$\nabla \times \left(\nabla \times \vec{E}\right) + \varepsilon_0 \mu_0 \frac{\partial^2 \left(\varepsilon \vec{E}\right)}{\partial t^2} = 0. \tag{7}$$

Assuming that the material effect on the electric field is slow compare to period of optical oscillation or laser pulse duration, i.e. assuming time independence of the permittivity of material, assuming that the material is weakly absorbing, and recalling that the speed of electro-magnetic wave in vacuum is $c = 1/\sqrt{\varepsilon_0 \mu_0}$ and index of refraction of material is $n = \sqrt{\varepsilon}$, we re-write equation (7) as

$$\nabla \times \left(\nabla \times \vec{E}\right) + \frac{n^2}{c^2} \frac{\partial^2 \vec{E}}{\partial t^2} = 0. \tag{8}$$

Now, using the identity $\nabla\times\left(\nabla\times\vec{A}\right)=\nabla\left(\nabla\bullet\vec{A}\right)-\Delta\vec{A}$ we re-write equation (8) in the following form

$$\Delta\vec{E}-\frac{n^2}{c^2}\frac{\partial^2\vec{E}}{\partial t^2}-\nabla\left(\nabla\bullet\vec{E}\right)=0\,. \tag{9}$$

From this point the derivations will significantly deviate from the procedure commonly performed in all the books and journal publication on nonlinear optics since we will be considering the permittivity of material and, consequently, the index of refraction as a coordinate dependent function. In contrast to our approach, customary considerations treat permittivity as a constant, zeroing the third term in the left hand side of equation (9) and reducing this equation to the commonly known form that contains only two first terms and is called the wave equation. As we show below, neglecting the third term is a significant mistake that leads to inadequate description of wave propagation in nonlinear media.

For the conditions of inhomogeneity of the properties of electrically neutral media, such as in case of propagation of short wavelength waves in ionosphere [13], equation (9) can be re-written in following form

$$\Delta\vec{E}-\frac{n^2}{c^2}\frac{\partial^2\vec{E}}{\partial t^2}+\nabla\left(\frac{\nabla\varepsilon\bullet\vec{E}}{\varepsilon}\right)=0\,, \tag{10}$$

where we used equations (1) and (5) from which it follows that $\nabla\varepsilon\bullet\vec{E}+\varepsilon\nabla\bullet\vec{E}=0$ and, therefore, $\nabla\bullet\vec{E}=-\frac{\nabla\varepsilon\bullet\vec{E}}{\varepsilon}$. Finally, recalling that, $\varepsilon=n^2$ and using convenient expression for the displacement field $\vec{D}=\varepsilon_0\varepsilon\vec{E}=\varepsilon_0(1+\chi)\vec{E}=\varepsilon_0\vec{E}+\vec{P}$, where $\chi$ is the electric susceptibility and $\vec{P}$ is the polarization density, we re-write equation (7) in following form

$$\Delta\vec{E}-\frac{1}{c^2}\frac{\partial^2\vec{E}}{\partial t^2}+2\nabla\left(\frac{\nabla n\bullet\vec{E}}{n}\right)=\mu_0\frac{\partial^2\vec{P}_L}{\partial t^2}+\mu_0\frac{\partial^2\vec{P}_{NL}}{\partial t^2}\,, \tag{11}$$

where the polarization density vector is represented by the sum of linear and nonlinear components denoted using subscripts "*L*" and "*NL*", respectively.

The solution of propagation equation expressed in form of either equation (10) or equation (11) can be expressed in terms of slowly varying amplitude function

$$\vec{E}(\vec{r},t)=\vec{A}(\vec{r})e^{i\left(\vec{k}(\vec{r})\bullet\vec{r}-\omega t\right)}+cc\,. \tag{12}$$

In this solution the vector amplitude of the electric field and the wave vector are coordinate dependent, i.e. expression (12) represents electric field of a non-planar wave.

Below we will demonstrate that, within paraxial approximation and considering propagation range in which variation of the spatial profile of laser beam irradiance due to the effect of nonlinear induced refraction is small and, thus, can be considered as perturbation, the propagation equations (10) or (11) can be straightforwardly modified into equation that, as proposed in [11,12], has a solution represented by the blend of solution of Helmholtz equation for propagation of a laser beam in a media with uniform and irradiance independent refractive index similar to one obtained by Kogelnik and Li [14] and a correction term that represents nonlinear field perturbation expressed in terms of paraxial ray-optics (eikonal equation) [15].

## 3. Unveiling faults of current theory for laser beam propagation in NL media

Now we will demonstrate that the current formulations of propagation equation in non-linear (self-induced inhomogeneity) media are based on two catastrophically flawed assumptions. First, in all theoretical works known to us it is assumed that the laser beam has plane wave front. Second, as mentioned above, the term responsible for refraction due to media inhomogeneity is disregarded, i.e. media is assumed as homogenous.

Under the first assumption, i.e. assumption of plane wave front, the solution of propagation equation is expressed in form of slowly varying amplitude function

$$\vec{E}(x,y,z,t) = A_x(x,y,z)e^{i(k_z z-\omega t)}\hat{x} + A_y(x,y,z)e^{i(k_z z-\omega t)}\hat{y} + cc\,. \qquad (13)$$

Note that here, in accordance to the plane wave assumption, the scalar product of vectors $\vec{k} \bullet \vec{r}$ from equation (12) is substituted with product of scalars $-k_z z$, where $k_z$ is the component of wave vector along the $z$ axis. The solution form (13) assumes that the $x$- and $y$-components of the wave vector and $z$-component of electric field are zero, i.e. beam propagates exactly along the $z$ axis.

According to the second assumption, the term describing refraction on gradient of refractive index is neglected. Then, assuming linear polarization and, since the requirement of slowly varying amplitude implies that,

$$\left|\frac{\partial^2 A}{\partial z^2}\right| << \left|k\frac{\partial A}{\partial z}\right|\,, \qquad (14)$$

the equation (10) can be reformulated in form retaining only the amplitude $A(x,y,z)$ of electric vector aligned along either x-axis of y axis

$$\Delta_T A + \left(\frac{\omega^2 n^2}{c^2} - k_z^2\right)A + 2ik_z\frac{\partial A}{\partial z} = 0\,. \qquad (15)$$

By definition, the wave vector $k = \frac{\omega n}{c}$, and since plane wave is assumed, the second term in equation (15) must be zero. However, all textbooks and scientific articles at this stage of consideration submit that, the wave front deviates from a plane. This, of course, contradicts to the initial assumption of strictly plane front; however, is necessary, as otherwise the self-focusing and self-trapping would not follow from equation (15). Thus, the second term of propagation equation (15) in all current models is assumed to be non-zero. We will show below that this manipulation has devastating consequence.

Another flaw of current theory of electromagnetic wave propagation in nonlinear media is that the third term in complete propagation equation (11) is omitted leading to commonly used equation (15). As far as we know a justification for such omission was never provided. It appears that, the origin of this obvious error can be traced to the original work [16]. In this work the mathematical consideration of the laser beam propagation in nonlinear medium started from equation in which the term describing refraction due to the medium inhomogeneity was omitted. Surprisingly, all the subsequent works, except for our recent publications [11,12], followed the path laid by [16] and labored on various mathematical treatments of equation that can be traced to the propagation equation deduced for homogenous media. Thus, the "foundational" propagation equation in nonlinear optics was deduced while disregarding media inhomogeneity (inherent, induced or self-induced), and has form of equation (15) missing the "refraction" term (see for example, equation 7.2.9 in [3]).

At this point one should wonder of how equation with term missing the refraction due to media inhomogeneity can describe self-focusing? The answer is hidden in the second term of equation (15). The trick with "non-zeroing" of the second term in equation (15) becomes handy and is crucial for constructing all nonlinear propagation effects out of this essentially irrelevant equation. Indeed, introducing nonlinearity of refractive index, $n = n_0 + n_2\left\langle |E|^2 \right\rangle_t$, equation (15) can be modified into a following form [17,18]

$$\frac{\partial^2 E}{\partial x^2} + 2ik\frac{\partial E}{\partial z} = -k^2\frac{n_2}{n_0}|E|^2 E\,, \qquad (16)$$

that is equivalent to the infamous (in the realm of nonlinear wave propagation) nonlinear Schrodinger's equation [19]

$$\frac{\partial^2 \psi}{\partial x^2} + i\frac{\partial \psi}{\partial z} = -\kappa|\psi|^2\psi\,. \qquad (17)$$

## 4. Revision of scientific principles - Επιστημονικές αρχή αξιολόγησης (epistomionic arxi-aksiology)

Let's revise the foundation of the current theory for propagation of a beam of electromagnetic wave in nonlinear medium. In this consideration we will use general form of 3D propagation equation (7) that follows directly from the Maxwell's equations. Following traditional approach, we will look for the solution of this equation in the following form of wave with varying amplitude

$$\begin{gathered}\vec{E} = E_x\hat{x} + E_y\hat{y} + E_z\hat{z} = \vec{A}(x,y,z)p(t)e^{i(k_x\hat{x}+k_y\hat{y}+k_z\hat{z}-\omega t)} = \\ [A_x(x,y,z)\hat{x} + A_y(x,y,z)\hat{y} + A_z(x,y,z)\hat{z}]p(t)e^{i(k_x\hat{x}+k_y\hat{y}+k_z\hat{z}-\omega t)},\end{gathered} \qquad (18)$$

where function *p(t)* is the dimensionless pulse shape such that time integral of this function from minus to plus infinity equals unity.

Let's modify now propagation equation (9) assuming that the pulse shape is slowly varying function compare to the period of wave oscillation and that the dielectric constant, ε, and therefore the refractive index, *n*, are time independent. Then, recalling that, $\varepsilon = n^2$, and substituting solution (18) into equation (9), neglecting time derivative of pulse shape *p(t)* as it is slow function, and eliminating exponent of phase, equation (9) can be transformed into equation for the amplitude:

$$\left(k_x^2+k_y^2+k_z^2-\frac{\omega^2 n^2}{c^2}\right)A_x\hat{x}+\left(k_x^2+k_y^2+k_z^2-\frac{\omega^2 n^2}{c^2}\right)A_y\hat{y}+\left(k_x^2+k_y^2+k_z^2-\frac{\omega^2 n^2}{c^2}\right)A_z\hat{z}+$$

$$\left[\begin{array}{l}\Delta A_x+2i\left(k_x\frac{\partial A_x}{\partial x}+k_y\frac{\partial A_x}{\partial y}+k_z\frac{\partial A_x}{\partial z}\right)-i\left(\frac{\partial k_x}{\partial x}+\frac{\partial k_y}{\partial y}+\frac{\partial k_z}{\partial z}\right)A_x+\\ 2i\left(x\frac{\partial k_x}{\partial x}\frac{\partial A_x}{\partial x}+y\frac{\partial k_y}{\partial y}\frac{\partial A_x}{\partial y}+z\frac{\partial k_z}{\partial z}\frac{\partial A_x}{\partial z}\right)-i\left(x\frac{\partial^2 k_x}{\partial x^2}+y\frac{\partial^2 k_y}{\partial y^2}+z\frac{\partial^2 k_z}{\partial z^2}\right)A_x+\\ 2\left(xk_x\frac{\partial k_x}{\partial x}+yk_y\frac{\partial k_y}{\partial y}+zk_z\frac{\partial k_z}{\partial z}\right)A_x+\left(x^2\left(\frac{\partial k_x}{\partial x}\right)^2+y^2\left(\frac{\partial k_y}{\partial y}\right)^2+z^2\left(\frac{\partial k_z}{\partial z}\right)^2\right)A_x\end{array}\right]\hat{x}+$$

$$\left[\begin{array}{l}\Delta A_y+2i\left(k_x\frac{\partial A_y}{\partial x}+k_y\frac{\partial A_y}{\partial y}+k_z\frac{\partial A_y}{\partial z}\right)-i\left(\frac{\partial k_x}{\partial x}+\frac{\partial k_y}{\partial y}+\frac{\partial k_z}{\partial z}\right)A_y+\\ 2i\left(x\frac{\partial k_x}{\partial x}\frac{\partial A_y}{\partial x}+y\frac{\partial k_y}{\partial y}\frac{\partial A_y}{\partial y}+z\frac{\partial k_z}{\partial z}\frac{\partial A_y}{\partial z}\right)-i\left(x\frac{\partial^2 k_x}{\partial x^2}+y\frac{\partial^2 k_y}{\partial y^2}+z\frac{\partial^2 k_z}{\partial z^2}\right)A_y+\\ 2\left(xk_x\frac{\partial k_x}{\partial x}+yk_y\frac{\partial k_y}{\partial y}+zk_z\frac{\partial k_z}{\partial z}\right)A_y+\left(x^2\left(\frac{\partial k_x}{\partial x}\right)^2+y^2\left(\frac{\partial k_y}{\partial y}\right)^2+z^2\left(\frac{\partial k_z}{\partial z}\right)^2\right)A_y\end{array}\right]\hat{y}+$$

$$\left[\begin{array}{l}\Delta A_z+2i\left(k_x\frac{\partial A_z}{\partial x}+k_y\frac{\partial A_z}{\partial y}+k_z\frac{\partial A_z}{\partial z}\right)-i\left(\frac{\partial k_x}{\partial x}+\frac{\partial k_y}{\partial y}+\frac{\partial k_z}{\partial z}\right)A_z+\\ 2i\left(x\frac{\partial k_x}{\partial x}\frac{\partial A_z}{\partial x}+y\frac{\partial k_y}{\partial y}\frac{\partial A_z}{\partial y}+z\frac{\partial k_z}{\partial z}\frac{\partial A_z}{\partial z}\right)-i\left(x\frac{\partial^2 k_x}{\partial x^2}+y\frac{\partial^2 k_y}{\partial y^2}+z\frac{\partial^2 k_z}{\partial z^2}\right)A_z+\\ 2\left(xk_x\frac{\partial k_x}{\partial x}+yk_y\frac{\partial k_y}{\partial y}+zk_z\frac{\partial k_z}{\partial z}\right)A_z+\left(x^2\left(\frac{\partial k_x}{\partial x}\right)^2+y^2\left(\frac{\partial k_y}{\partial y}\right)^2+z^2\left(\frac{\partial k_z}{\partial z}\right)^2\right)A_z\end{array}\right]\hat{z}- \quad . (19)$$

$$2\left[\nabla\left(\frac{\nabla n}{n}\bullet(\vec{A}e^{i(k_x x+k_y y+k_z z)})\right)\right]e^{-i(k_x x+k_y y+k_z z)}=0$$

Note, that first three terms in the left hand side of equation (19) are null since, by definition, $k_x^2+k_y^2+k_z^2-\frac{\omega^2 n^2}{c^2}=0$ and, as in previous deduction of equation (10), we used equation $\nabla\bullet\vec{E}=-\frac{\vec{E}\nabla\varepsilon}{\varepsilon}$ in order to formulate right hand side of equation (19). Thus, the propagation equation in nonlinear media has the following form

$$\sum_{m=x,y,z}\left[\begin{array}{l}\Delta A_m+2i\left(k_m\frac{\partial A_m}{\partial x}+k_y\frac{\partial A_m}{\partial y}+k_z\frac{\partial A_m}{\partial z}\right)-i\left(\frac{\partial k_x}{\partial x}+\frac{\partial k_y}{\partial y}+\frac{\partial k_z}{\partial z}\right)A_m+\\ 2i\left(x\frac{\partial k_x}{\partial x}\frac{\partial A_m}{\partial x}+y\frac{\partial k_y}{\partial y}\frac{\partial A_m}{\partial y}+z\frac{\partial k_z}{\partial z}\frac{\partial A_m}{\partial z}\right)-i\left(x\frac{\partial^2 k_x}{\partial x^2}+y\frac{\partial^2 k_y}{\partial y^2}+z\frac{\partial^2 k_z}{\partial z^2}\right)A_m+\\ 2\left(xk_x\frac{\partial k_x}{\partial x}+yk_y\frac{\partial k_y}{\partial y}+zk_z\frac{\partial k_z}{\partial z}\right)A_m+\left(x^2\left(\frac{\partial k_x}{\partial x}\right)^2+y^2\left(\frac{\partial k_y}{\partial y}\right)^2+z^2\left(\frac{\partial k_z}{\partial z}\right)^2\right)A_m\end{array}\right]\hat{m}-$$

$$2\left[\nabla\left(\frac{\nabla n}{n}\bullet(\vec{A}e^{i(k_x x+k_y y+k_z z)})\right)\right]e^{-i(k_x x+k_y y+k_z z)}=0. \quad (20)$$

This equation is cardinally different from the currently used equations (15), (16) and (17) in both aspects of physics and mathematics. From point of view of a physicist, Equation (21) straightforwardly shows that beam refraction is produced by self-induced inhomogeneity of refractive index (that can be result of Kerr effect, material ionization or else). Also, a physicist will find appealing that the presented theory expressed by equation (20) satisfies correspondence principle since, as we will demonstrate below, it contains geometric optics and equation (20) transforms into ray optics equation under assumption of infinitely small wavelength. In contrast, equations (15), (16) and (17) don't lead to the geometric optics in limiting case of infinitely small wavelength and, thus don't satisfy the correspondence principle.

From the point of view of a mathematician new equation of propagation (20) dramatically differs from the current propagation equation in one very peculiar aspect – it has no self-similar solution. In contrast, equation (16) re-written as nonlinear Schrodinger's equation (17) has self-similar solution or, so called, soliton solution that serves as foundation for prediction of laser beam self-trapping and all the current "filament" theories that predict mind boggling lengths of self-trapped laser "filament".

## 5. Modification of new propagation equation for case of paraxial laser beam propagation in nonlinear media at distance shorter of comparable to Ryleigh length

Let's explore the complete 3-D propagation equation (20) within paraxial beam approximation while considering propagation range in which perturbation of the spatial profile of laser beam irradiance by the nonlinear induced refraction is negligible. The latter condition is realized within the range of several Ryleigh lengths for a focused laser beam with pulse energy that is below certain value (see detailed discussion in [11,12]).

Here we will demonstrate that the propagation equation (20) leads, as proposed in [11,12], to the solution represented by the blend of solution of Helmholz equation for propagation of a laser beam in a media with uniform and irradiance independent refractive index similar to one obtained by Kogelnik and Li [14] and a correction term that represents nonlinear field perturbation expressed in terms of paraxial ray-optics (eikonal) equation [15].

Assuming paraxial beam propagation it is easy to see that one can neglect in propagation equation (20) the terms containing *x*- and *y*- components of the wave vector and the terms containing *z*- component of the electric field amplitude as well as their derivatives (see the schematic of the paraxial beam propagation in Figure 1).

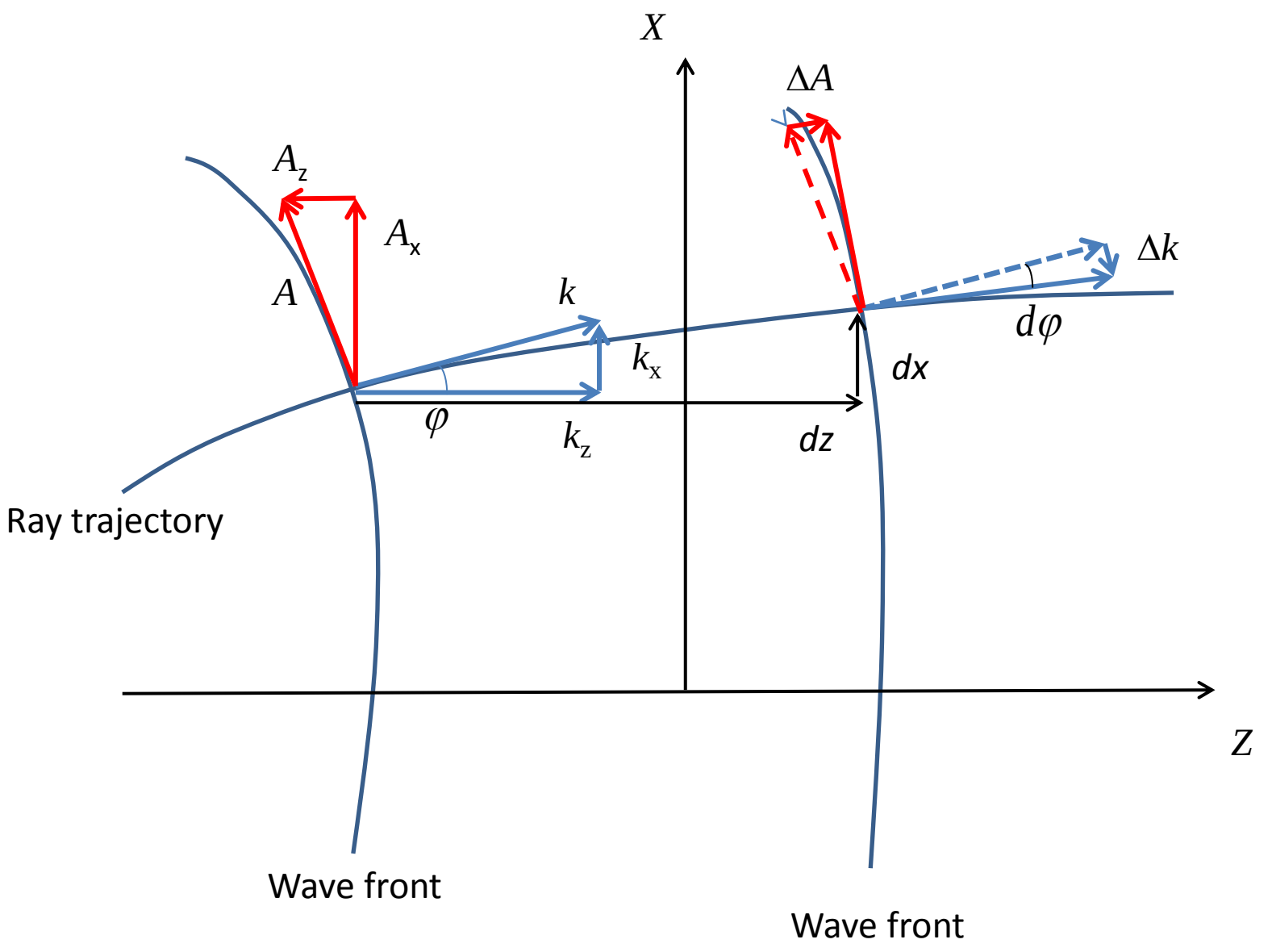

Figure 1. Schematic of evolution of electric field amplitude and wave vector during laser beam propagation.

Assuming linear polarization in the *x*-z plane and neglecting smaller terms as described above allows following simplification of propagation equation (20):

$$\begin{bmatrix} \left(\frac{\partial^2 A_x}{\partial x^2}+\frac{\partial^2 A_x}{\partial y^2}\right)+2i\left(k_x\frac{\partial A_x}{\partial x}+k_z\frac{\partial A_x}{\partial z}\right)-i\left(\frac{\partial k_x}{\partial x}+\frac{\partial k_z}{\partial z}\right)A_x+ \\ 2i\left(x\frac{\partial k_x}{\partial x}\frac{\partial A_x}{\partial x}+z\frac{\partial k_z}{\partial z}\frac{\partial A_x}{\partial z}\right)-i\left(x\frac{\partial^2 k_x}{\partial x^2}+z\frac{\partial^2 k_z}{\partial z^2}\right)A_x+ \\ 2\left(xk_x\frac{\partial k_x}{\partial x}+zk_z\frac{\partial k_z}{\partial z}\right)A_x+\left(x^2\left(\frac{\partial k_x}{\partial x}\right)^2+z^2\left(\frac{\partial k_z}{\partial z}\right)^2\right)A_x \end{bmatrix}\hat{x}+$$

$$\begin{bmatrix} \left(\frac{\partial^2 A_z}{\partial x^2}+\frac{\partial^2 A_z}{\partial y^2}\right)+2i\left(k_x\frac{\partial A_z}{\partial x}+k_z\frac{\partial A_z}{\partial z}\right)-i\left(\frac{\partial k_x}{\partial x}+\frac{\partial k_z}{\partial z}\right)A_z+ \\ 2i\left(x\frac{\partial k_x}{\partial x}\frac{\partial A_z}{\partial x}+z\frac{\partial k_z}{\partial z}\frac{\partial A_z}{\partial z}\right)-i\left(x\frac{\partial^2 k_x}{\partial x^2}+z\frac{\partial^2 k_z}{\partial z^2}\right)A_z+ \\ 2\left(xk_x\frac{\partial k_x}{\partial x}+zk_z\frac{\partial k_z}{\partial z}\right)A_z+\left(x^2\left(\frac{\partial k_x}{\partial x}\right)^2+z^2\left(\frac{\partial k_z}{\partial z}\right)^2\right)A_z \end{bmatrix}\hat{z}-$$

$$2\left[\nabla\left(\frac{\nabla n}{n}\bullet(\vec{A}e^{i(k_x x+k_y y+k_z z)})\right)\right]e^{-i(k_x x+k_y y+k_z z)}\approx$$

$$\left[\left(\frac{\partial^2 A_x}{\partial x^2}+\frac{\partial^2 A_x}{\partial y^2}\right)+2ik_z\frac{\partial A_x}{\partial z}\right]\hat{x}+\left[2ik_z\frac{\partial A_z}{\partial z}\right]\hat{z}-2\left[\nabla\left(\frac{\nabla n}{n}\bullet(\vec{A}e^{i(k_x x+k_y y+k_z z)})\right)\right]e^{-i(k_x x+k_y y+k_z z)}=0 \quad . (21)$$

Then, from simplified propagation equation (21) we can extract two equations for *x* coordinate

$$\left(\frac{\partial^2 A_x}{\partial x^2}+\frac{\partial^2 A_x}{\partial y^2}\right)+2ik_z\frac{\partial A_x}{\partial z}+2\left[\nabla_x\left(\frac{\nabla n}{n}\cdot\vec{A}e^{i(k_x x+k_y y+k_z z)}\right)\right]e^{-i(k_x x+k_y y+k_z z)}=0\,, \qquad (22)$$

and for *z* coordinate

$$2ik_z\frac{\partial A_z}{\partial z}+2\left[\nabla_z\left(\frac{\nabla n}{n}\bullet(\vec{A}e^{i(k_x x+k_y y+k_z z)})\right)\right]e^{-i(k_x x+k_y y+k_z z)}=0\,. \qquad (23)$$

The third term in the left-hand side of equation (22) for *x*- (transverse) component of the amplitude of the electric field is small compared to the first two terms and for the short propagation distances it can be neglected. Then, equation (22) acquires form similar to the equation obtained for diffraction dominated laser beam propagation in empty resonators [14]. The solutions of equation (22) while neglecting third term can be found in the classical article of Kogelnik and Li [14] and according to this work, the fundamental mode of the solution is represented by the field that has Gaussian distribution of amplitude in the radial direction with the beam width that changes along the *z*-axis and has spherical shape of the wave front with the radius that is also a function of *z*. Of course, for far field propagation the third term in the left-hand side of equation (22) must be accounted for since the relatively small deviations of the amplitude of electric field and the shape of the wave front should accumulate while propagation at long distances resulting in significant modification of both beam intensity distribution and the wave front shape.

Both terms in the left-hand side of equation (23) have similar magnitude. Now, let's demonstrate that the solution of equation (23) describes perturbation of the wave front of the "carrier" filed given by the approximate solution of equation (22) described above.

$$\begin{aligned}2ik_z\frac{\partial A_z}{\partial z}\hat{z}&=-2\left[\nabla_z\left(\frac{\nabla n}{n}\bullet(\vec{A}e^{i(k_x x+k_y y+k_z z)})\right)\right]e^{-i(k_x x+k_y y+k_z z)}=\\&=-2e^{-i(k_x x+k_y y+k_z z)}\frac{\partial}{\partial z}\left(\frac{1}{n}\frac{\partial n}{\partial x}A_x e^{i(k_x x+k_y y+k_z z)}\right)\hat{z}=\\&-2\left[\left(\frac{1}{n}\frac{\partial^2 n}{\partial z\partial x}-\frac{(\partial n/\partial z)(\partial n/\partial x)}{n^2}\right)A_x+\frac{1}{n}\frac{\partial n}{\partial x}\left(\frac{\partial A_x}{\partial z}+ik_z A_x\right)\right]\hat{z}\approx-2ik_z\frac{1}{n}\frac{\partial n}{\partial x}A_x\hat{z}\end{aligned}\,, \qquad (24)$$

or

$$\frac{\partial A_z}{\partial z}\approx-\frac{1}{n}\frac{\partial n}{\partial x}A_x\,. \qquad (25)$$

Finally, recalling that in paraxial approximation change of the angle between the wave vector and $z$-axis, $d\varphi$, equals to the ratio of the $x$ and $z$ components of the wave vector or to the ratio of the $z$ and $x$ components of the electric field amplitude, equation (25) can be re-written as

$$d\varphi = \frac{-\delta A_z}{A_x} \approx \frac{1}{n}\frac{\partial n}{\partial x} dz\,, \tag{26}$$

which is the eikonal equation [15]. We use integration of this equation (see, for example, equation (7) in [12]) in our previous works [11,12] in order to compute radial distribution of the angle of wave vector after the propagation of focused laser beam through caustic under conditions when focusing due to Kerr effect and defocusing due to ionization takes place (see, for example, equation (8) in [12]).

At the same time, the local projections of the wave vector are uniquely determined from the system of equations:

$$\begin{aligned} & k_x^2 + k_z^2 = \frac{\omega^2 n^2}{c^2} \\ & \frac{k_x}{k_z} = \mathrm{tg}\varphi \end{aligned}\,, \tag{27}$$

with the solution:

$$k_z(x,z) = \frac{\omega n(x,z)}{c\left(1 + \mathrm{tg}\varphi(x,z)\right)^{1/2}}\,,\quad k_x(x,z) = \frac{\omega n(x,z)\mathrm{tg}\varphi(x,z)}{c\left(1 + \mathrm{tg}\varphi(x,z)\right)^{1/2}}\,. \tag{28}$$

Finally, complete equation (20) for electro-magnetic wave propagation written, without affecting its generality, for the linear polarization in $x$-$z$ plane (or instead its simplified equivalent (21) deduced for paraxial and near field propagation) in combination with the equations (26-28) and with addition of equation describing inhomogeneity of the index of refraction (either self-induced due to the Kerr effect and ionization or externally induced due to thermal effect, large scale turbulence, etc. or inherent due to spatially variable material properties) represent closed system of equations that provides unique solution describing electro-magnetic wave propagation in inhomogeneous medium.

## Concluding remarks

Because of obvious reasons, it is usual practice in majority of non-specialized educational courses and textbooks to ignore contribution of the term containing $\nabla\varepsilon$ and describe electro-magnetic wave propagation using equation deduced for uniform media. Unfortunately, without much consideration, this propagation equation was adopted in nonlinear optics and after some peculiar modifications it took form of equations (15, 16). Then, relatively recent trend mandated further modification of this simplified equations in order to acquire appearance similar to nonlinear Schrodinger's equation (17). We will leave for the future debates obviously devious nature of such misnomer, as electro-magnetic wave propagation still remains in the realm of classical physics. Rather, we would like to reinforce that omitting the third term in the left-hand side of the above derived complete propagation equation (9) leads to cascading failures of the current models because of inadequate description of the physics involved. Indeed, it is rather dubious to disregard the inhomogeneity of optical properties of media (self-induced, externally induced or inherent) on the stage of deduction of the propagation equation and then to re-introduce into the obtained simplified equation nonlinear dependence of index of refraction on the laser irradiance. Trivial estimates show non-negligible contribution of the term containing $\nabla\varepsilon$ in complete propagation equation (10) that, in particular, describes laser beam propagation in nonlinear media. Therefore, ignoring this term leads to false predictions. An example of such false prediction is prediction of waveguide like propagation of laser beam in nonlinear media [16] that lead to the development of fascinating concepts of optical soliton and laser beam filamentation that recently produced a flurry of extensive research. It is easy to see that for a laser beam with intensity distribution that is near-Gaussian a possible solution of incorrectly abbreviated propagation equation, such as (15,16,17), indeed, has a self-similar form. Currently accepted interpretation of this self-similarity is that the divergence due to diffraction and media ionization is compensated by self-focusing [2-4]. One of the results, following from the self-similarity of the solution of this inadequate propagation equation (15,16,17) is a captivating (however, contradicting to experimental observations) prediction that kilometers long transmission of the laser beam can be achieved in atmosphere without beam divergence [6] resulting in kilometers long filaments of ionized air.

Our work demonstrates that the solution of complete equation (10) that adequately describes laser beam propagation in nonlinear media does not have self-similar form. As a demonstration we solved complete propagation equation for the conditions when input from the nonlinear refraction can be treated as perturbation of the solution of the linear Helmholtz equation describing propagation of focused laser beam [11,12]. This solution demonstrated that laser beam divergence is affected by Kerr self-focusing and plasma defocusing differently in different radial locations of the laser beam and in different times during the laser pulse, i.e. self-similar beam propagation does not occur.

Another catastrophic flaw of the customary approach in which equation (17) or its derivatives are utilized for modeling of diverging or converging laser beam propagation in nonlinear media is that the equation (17) is derived under assumption of plane wave front. Consequently, it is inapplicable for description of non-planar wave fronts. However, in a dubious manner similar to the above illustrated re-insertion of the nonlinear effect into equation (17), the complex non-planar wave propagation is described in all current theoretical models using plane wave propagation equation (17) or its derivatives. In particular, it is worth mentioning that the Poynting vector maintains its direction in the approximation of a plane wave front. In contrast, for a converging of diverging laser beam the local direction of the Poynting vector varies as function of the distance from the axis. Our previous eikonal-paraxial model [11,12], despite its simplicity, correctly reflects this dependence.

The effect of inhomogeneity of the dielectric constant on the electro-magnetic wave propagation is known in several relatively narrowly specialized fields. However, overwhelming majority of optics textbooks, including all textbooks on nonlinear optics, as well as all nonlinear optics research publications present and operate with simplified propagation equation failing to disclose general theoretical concept of wave propagation in inhomogeneous media. So far, the practical application of this concept was limited to the theory of radio wave propagation in ionosphere (see reference [13]) and electro-magnetic and acoustic wave propagation in stratified media [20], such as radar propagation in atmospheric boundary layers [21]. It is reasonable to suggest that obscurity of the theory of wave propagation in inhomogeneous media is the reason of why nonlinear optics theorists failed to recognize importance of self-induced media inhomogeneity and neglected formulation of correct propagation equation. Instead, all the effort was concentrated on including nonlinearity into the abbreviated propagation equation formulated under assumptions of media homogeneity and planar wave front and solving this modified abbreviated equation.

In conclusion we summarize the contribution of this work to the field of nonlinear optics: 1) realization that gradient of dielectric constant always provides non-negligible contribution in the propagation equation of a laser beam with realistic beam profile because the characteristic length of change of irradiance is comparable to the "beam size" (for any reasonable definition of this physical property), and 2) development of the method described in [11,12] in which we integrated diffractive and geometrical optics by blending solution of linear Maxwell's equation and a correction term that represents nonlinear field perturbation expressed as solution of paraxial ray-optics (eikonal) equation that opened an elegant way for numerical computation of the ray trajectories (avoiding singularities) as the focused laser beam propagates in a nonlinear and ionized media through its caustic (the area near the focal plane that extends several Rayleigh lengths).

The realm of nonlinear optics that deals with the laser beam propagation benefited from multitude of experimental works and significant experience was accumulated in solving complex mathematical problems; however, it seems that substantial improvement of the theoretical part of nonlinear optics is needed and revision of the fundamentals of the theoretical model provided in our work can re-vitalize and substantially advance this field.

## Acknowledgments

One of the authors (VS) would like to thank DARPA and, particularly, Dr. R.B. for support of a project that led to realization of inconsistencies of the widely accepted theoretical model of laser filamentation. Also, the authors would like to thank Kelly Semak for her help and encouragement in the process of work on this manuscript.